\documentclass[12pt]{article}

\usepackage{epsfig}

\makeatletter
\def\textfraction{0.2}
\@textmin = \textfraction\textheight
\def\fps@figure{tbp} % Changed RBSz default placement for figures
\def\fps@table{tbp} % Changed RBSz default placement for tables
\makeatother

\def\dirc{{\sc Dirc}}
\def\Babar{\mbox{\sl B\hspace{-0.4em} {\scriptsize\sl A}\hspace{-0.4em} B\hspace{-0.4em} {\scriptsize\sl A\hspace{-0.1em}R}}}
\def\babar{\mbox{\sl B\hspace{-0.4em} {\scriptsize\sl A}\hspace{-0.4em} B\hspace{-0.4em} {\scriptsize\sl A\hspace{-0.1em}R}}}

\begin{document}

%%%%% The following lines create the SLAC Pub Title Page
%%
\thispagestyle{empty}
\renewcommand{\thefootnote}{\fnsymbol{footnote}}

%%%%% Substitute your Pub number, month and year in the following:
%%
\begin{flushright}
{\small
SLAC--PUB--8590\\
August 2000\\}
\end{flushright}

\vspace{.8cm}

%%%%% Title and Author Information:
%%
\begin{center}
{\bf\large DIRC, the Particle Identification System for 
BABAR\footnote{
Work supported by the Department of Energy under contracts
DE-AC03-76SF00515 (SLAC), DE-AC03-76SF00098 (LBNL), DE-AM03-76SF0010
(UCSB), and DE-FG03-93ER40788 
(CSU); the National Science Foundation
grant PHY-95-11999 (Cincinnati). }}

\vspace{1cm}

{\large Jochen Schwiening}\\[2mm]
Stanford Linear Accelerator Center, Stanford University,
Stanford, CA  94309\\

\vspace{1cm}

{\large Representing the BABAR-DIRC Collaboration~$^\ddagger$}
\end{center}

\vfill

\begin{center}
{\bf\large   
Abstract }
\end{center}

\begin{quote}
The \dirc , a novel type of Cherenkov ring imaging device, 
is the primary hadronic particle identification
system for the \Babar\ detector at the asymmetric B-factory, 
{\sc Pep-II} at SLAC.
%It is based on total internal reflection and uses long, rectangular bars
%made from synthetic fused silica as Cherenkov radiators and light guides.
\babar\ began taking data with colliding beams mode in late spring 1999.
This paper describes the performance of the \dirc\ during the first 16
months of operation.
\end{quote}

\vfill

%%%%%%%%%%%%%%%
%% Choose"Presented at," "Contributed to" for conference papers
%% or "Submitted to" for journal papers
%%%%%%%%%%%%%%%
\begin{center} 
{\it Invited talk presented at} \\
{\it 30th International Conference On High-Energy Physics (ICHEP 2000)}\\
{\it Osaka, Japan}\\
{\it Jul 27 - Aug 2, 2000} \\
\end{center}

\newpage
\begin{center} 
\begin{large} 
$^\ddagger$ The BABAR-DIRC Collaboration \\[1cm]

I.~Adam,$^1$
R.~Aleksan,$^2$
D.~Aston,$^1$ 
M.~Benkebil,$^4$ 
D.~Bernard,$^5$
G.~Bonneaud,$^5$
F.~Brochard,$^5$
D.N.~Brown,$^6$
P.~Bourgeois,$^2$
J.~Chauveau,$^3$
J.~Cohen-Tanugi,$^5$
M.~Convery,$^1$
G.~de Domenico,$^2$
A.~de Lesquen,$^2$
S.~Emery,$^2$
S.~Ferrag,$^5$
A.~Gaidot,$^2$
T.~Geld,$^9$
G.~Hamel de Monchenault,$^{2}$
C.~Hast,$^4$
A. Hoecker,$^4$
R.W.~Kadel,$^5$
J.~Kadyk,$^5$
H.~Lacker,$^4$
G.W.~London,$^2$ 
A.~Lu,$^7$
A.-M.~Lutz,$^4$
G.~Lynch,$^6$
G.~Mancinelli,$^9$
F.~Martinez-Vidal,$^3$
N.~Mayer,$^2$
B.T.~Meadows,$^9$
D.~Muller,$^1$
S.~Plaszczynski,$^4$
M.~Pripstein,$^6$ 
B.N.~Ratcliff,$^1$
L.~Roos,$^3$
E.~Roussot,$^5$
M.-H.~Schune,$^4$
J.~Schwiening,$^1$
V.~Shelkov,$^6$
M.D.~Sokoloff,$^9$
S.~Spanier,$^1$
J.~Stark,$^3$
A.V.~Telnov,$^6$
Ch.~Thiebaux,$^5$
G.~Vasileiadis,$^5$
G.~Vasseur,$^2$
J.~Va'vra,$^1$
M.~Verderi,$^5$
W.A.~Wenzel,$^6$
R.J.~Wilson,$^8$
G.~Wormser,$^4$
Ch.~Y{\'e}che,$^2$
S.~Yellin,$^7$
M.~Zito.$^2$ \\
\vskip 1cm

\end{large}
$^1$Stanford Linear Accelerator Center, Stanford, CA~94309, USA. \\
$^2$CEA, DAPNIA, CE-Saclay, F-91191, Gif-sur-Yvette Cedex, France. \\
$^3$LPNHE des Universit{\'e}s Paris 6 et Paris 7, Tour 33, Bc 200, 4 Place
Jussieu, F-75252, Paris, Cedex 05, France. \\
$^4$LAL Orsay, Universite Paris Sud, Batiment 200, F-91405 Orsay Cedex,
France. \\
$^5$LPNHE de l'Ecole Polytechnique, Route de Saclay, F-91128 Palaiseau
Cedex, France. \\ 
$^6$Lawrence Berkeley National Laboratory, One Cyclotron Road, Berkeley, CA
94720, USA. \\
$^7$Dept.  of Physics, University of California, Santa Barbara,
	CA~93106, USA. \\
$^8$Dept.  of Physics, Colorado State University, Fort Collins,
	CO~80523, USA. \\
$^9$Dept. of Physics, University of Cincinnati, Cincinnati,
	OH~45221, USA.

\end{center}
\newpage
%%
%%%%% End of title page

%%%%% Following are the commands to create the rest of the SLAC Pub.
%%
%%%%% The next two lines change the line spacing to doublespace,
%%      if you should need to do that.
%%
%\renewcommand{\baselinestretch}{2}
%\normalsize

%%%%% Your paper starts here:
%%

%% To get page numbers in the rest of the paper:
%
\pagestyle{plain}

The primary physics goal of the \Babar\ experiment~\cite{three} 
at the SLAC {\sc Pep-II} asymmetric $e^+e^-$ collider is to 
study CP violation in the $B^0$ meson system produced in $\Upsilon (4S)$
decays.
At the $\Upsilon (4S)$, {\sc Pep-II} collides 9~GeV electrons on
3.1~GeV positrons at
$\beta\gamma (lab) = 0.56$. 
The study of CP-violation in hadronic final states of the B meson
system requires the ability to tag the flavor of one of the $B$ mesons
via the cascade decay $b\rightarrow c \rightarrow s$, while fully
reconstructing the final state of the other B.
% over a large region of solid angle and momentum.
The momenta of the kaons used for flavor tagging extend up to about
2~GeV/c, with most below 1~GeV/c.
On the other hand, pions from the rare two-body decays $B^0\rightarrow
\pi^-\pi^+ (K^- \pi^+)$ must be well-separated from kaons, and have
momenta between 1.5 and 4.5~GeV/c 
where high momentum tracks are srongly correlated with 
forward polar angles due to the c.m. system boost.
Since the \Babar\ inner drift chamber tracker can provide $\pi/K$
separation up to approximately 700~MeV/c, an additional dedicated particle
identification system is required that must perform well over the
range of 700~MeV/c to about 4~GeV/c.

The system being used in \Babar\ is a novel type of ring imaging
Cherenkov detector, %based on total internal reflection, 
called the
\dirc\ ~\cite{princip} (Detection of Internally Reflected Cherenkov
light), which 
has been described in detail elsewhere~\cite{detector}.
Briefly, it uses 4.9~m long, rectangular bars made from synthetic fused silica 
% (Spectrosil\cite{six}) 
as Cherenkov radiator and light guide.
A charged particle with velocity $v$, traversing the fused silica bar
with index of refraction $n \ (\sim 1.473)$, generates a cone of
Cherenkov photons of half-angle $\theta_c$ with respect to the
particle direction, where $\cos \theta_c = 1/\beta n \ (\beta = v/c, 
\ c = $ velocity of light).
For particles with $\beta \approx 1$, some photons always lie
within the total internal reflection limit, and are transported 
efficiently to either one or both ends of the bar, depending on the
particle incident angle.
Since the bar has a rectangular cross section and is made to optical
precision, the magnitude of the Cherenkov angle is conserved during
the reflection at the radiator bar surfaces. 
% except for left-right/up-down ambiguities due to the
%
The photons are imaged via ``pin-hole" focussing by expanding through
a standoff region filled with 6000 litres of purified water 
onto an array of 10752 densely packed photomultiplier tubes
placed at a distance of about 1.2 m from the bar end.
Imaging in the \Babar\ \dirc\ occurs in three dimensions,
by recording the location and the time at which a given PMT is hit. 
The expected single photon Cherenkov angle resolution is 
about 9~mrad, dominated by a geometric term that is due to the sizes of
bars, PMTs and the expansion region, and a chromatic term from the
photon production. 
The accuracy of the time measurement is limited by the intrinsic
1.5~ns transit time spread of the PMTs.

In the absence of correlated systematic errors, the resolution 
($\sigma_{C,track}$) on the track Cherenkov angle 
should scale as
\begin{equation}
	\sigma_{C,track} = \sigma_{C,\gamma}/\sqrt{N_{pe}} \ ,
\end{equation}

\noindent
where $\sigma_{C,\gamma}$ is the single photon Cherenkov angle
resolution, and $N_{pe}$ is the number of photons detected.
The average single photon resolution obtained for photoelectrons from
di-muon events, $e^+e^-\rightarrow\mu^+\mu^-$, is 10.2~mrad, about
10\% worse than the expected value.
%There is a broad background of less than 10\%\
%relative height under the peak, that originates mostly from 
%track-associated sources. 
%The precise nature of these sources remain under investigation.
The time resolution obtained is 1.7~ns, close to the single-photon
resolution of the PMTs.  

\begin{figure}
\begin{center}
\epsfig{file=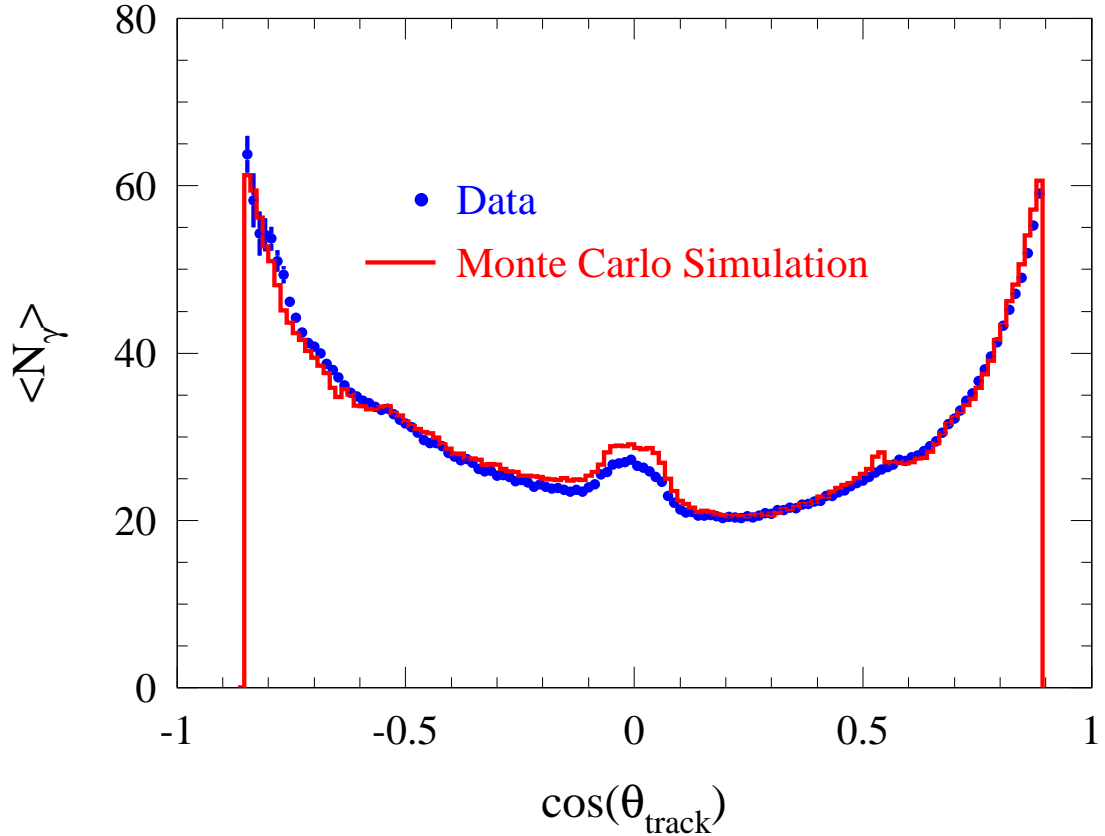,width=0.9\textwidth}
\end{center}
\label{fig:nphot}
\vspace*{-10mm}
\caption{
Number of detected photoelectrons {\it vs.} track dip angle 
$\theta_{\mathrm{track}}$ for di-muon events.}
\end{figure}
 
The number of photoelectrons per track, shown in Fig.~1, %%js
varies from a minimum of about 20 for small dip angles at the 
center of the barrel to well over 50 at large dip angles.
This is in good agreement with the value expected from the Monte Carlo
simulation at all angles.
This spectrum also demonstrates a very useful feature of the \dirc\ in the
\babar\ environment -- the performance improves in the 
forward direction, as is needed to cope with the angle-momentum
correlation of particles from the boost.

\begin{figure}
\begin{center}
\epsfig{file=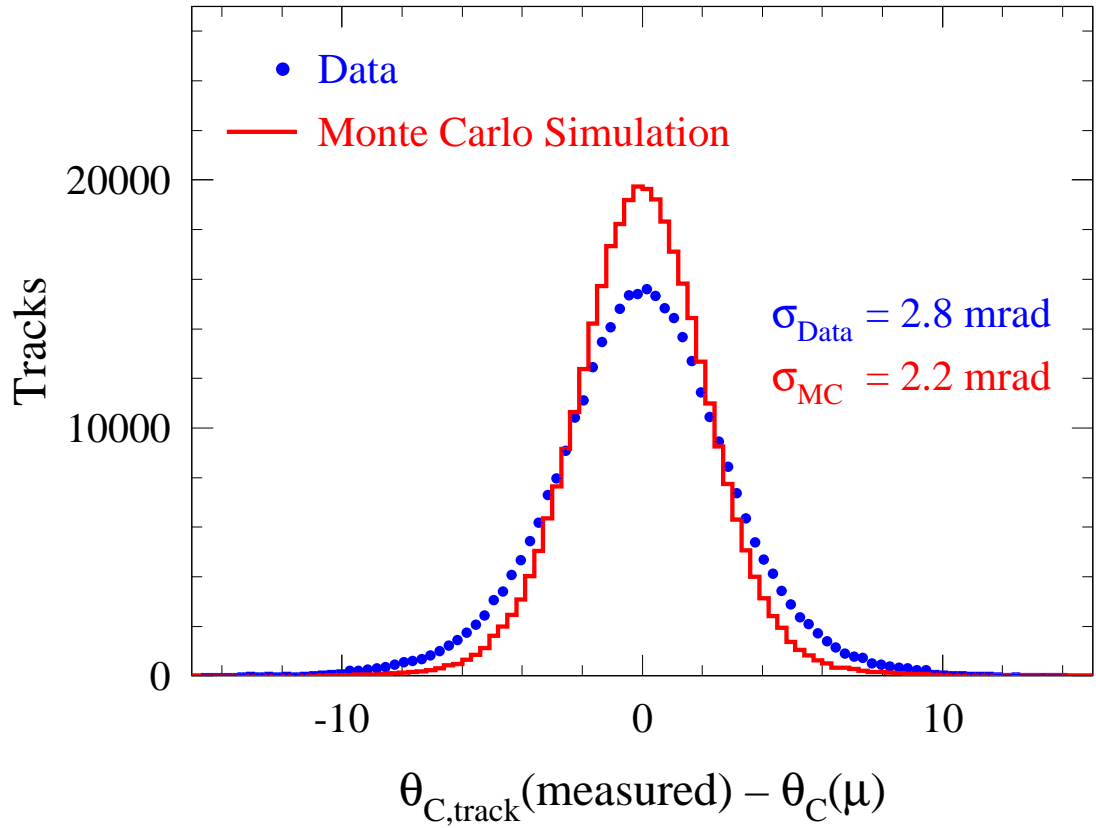,width=0.9\textwidth}
\end{center}
\label{fig:trackres}
\vspace*{-10mm}
\caption{
Resolution of the reconstructed Cherenkov polar angle per track for
di-muon events.}
\end{figure}
 
With the present alignment, the typical track Cherenkov angle
resolution for di-muon events is shown in Fig.~2 to be  %%js
2.8 mrad.
This is about 25\% worse than the 2.2~mrad expected from simulation.
From the measured single track resolution resolution $vs.$ momentum in
di-muon events and the difference between the expected Cherenkov
angles of charged pions and kaons, the pion-kaon separation power of
the \dirc\ can be inferred.
The present separation between kaons and pions at 3 GeV/c is about 3.8
$\sigma$, approximately 10\%\ worse than predicted by the Monte Carlo
simulation, and is expected to improve with advances in tracking and
detector alignment.

In summary, the \dirc\ was successfully commissioned, attained
performance rather close to that expected from Monte Carlo, and has
played a significant role in almost all \babar\ analyses presented at
this conference. 
The \dirc\ has been robust and stable and, almost 2 years after
installation, about 99.7\% of all PMTs and electronic channels are still
operating with nominal performance.

%\section{Conclusions}
%
%The \dirc\ is a novel ring-imaging Cherenkov detector that is
%well-matched to the hadronic particle identification requirements of
%\Babar .  
%The \dirc\ has run with partial azimuthal radiator coverage for all
%collision data taking since {\sc Pep-II} start-up in May 1999, and has
%been fully completed and operational since October 1999.  
%It runs very reliably, and is stable and easy to operate.  
%Initial performance obtained is already rather close to that simulated
%by the Monte Carlo. 
%Alignment and further code developments are underway which are
%expected to further improve performance soon.

\end{document}